\documentclass{desyproc}
\usepackage{color}
\usepackage[merge,numbers,compress]{natbib}
\usepackage{hyperref}

\definecolor{ourpurple}{RGB}{145,0,140}
\definecolor{darkorange}{RGB}{225,100,0}
\definecolor{darkgreen}{RGB}{0,170,0}

\hypersetup{
	colorlinks=false,        
	linkcolor=black,          
	citecolor=black,        
	filecolor=black,      
	urlcolor=black           
}

\begin{document}
\title{Dark Matter at the LHC and IceCube -- \\ a Simplified Model Interpretation}

\author{{\slshape Jan Heisig, Mathieu Pellen}\\[1ex]
Institute for Theoretical Particle Physics and Cosmology, RWTH Aachen University,
Germany\\}

\contribID{heisig\_jan}

\confID{11832}  
\desyproc{DESY-PROC-2015-02}
\acronym{Patras 2015} 
\doi  

\maketitle

\begin{abstract}
We present an interpretation of searches for Dark Matter in a simplified model approach. 
Considering Majorana fermion Dark Matter and a neutral vector mediator with axial-vector 
interactions we explore mono-jet searches at the LHC and searches for neutrinos from Dark 
Matter annihilation in the Sun at IceCube and place new limits on model parameter space.
Further, we compare the simplified model with its effective field theory approximation 
and discuss the validity of the latter one.
\end{abstract}

\section{Introduction}

Weakly interacting massive particles (WIMPs) are popular candidates to account 
for Dark Matter (DM) in the universe. There are different  ways to explore the landscape 
of DM models from a phenomenological perspective. One is the use of effective operators 
describing the interactions between the standard model (SM) and the WIMP in the 
framework of effective field theory (EFT). Another approach is to use simplified models
containing a minimal set of new particles that allow to describe the phenomenology of 
DM via renormalizable interactions. Although the EFT framework has been successfully 
used for the description of DM interactions at rather low energies -- like the interpretation 
of direct detection experiments -- it has been pointed out that EFT may break down when 
probing dark matter production at the 
LHC~\cite{Buchmueller:2013dya,Busoni:2014sya,Buchmueller:2014yoa}.

In this article we consider a model that extends the SM by a Majorana fermion DM, 
$\chi$, and a vector mediator, $V$, which couples to the DM and the SM quarks 
with axial-vector interactions, with couplings $g_\chi$ and $g_q$, respectively.
For such a model, LHC searches are expected to be more sensitive than direct 
detection experiments as the model does not provide any contribution to 
spin-independent WIMP-nucleon scattering. We present limits on the parameters 
space of this model from mono-jet searches at the 
LHC~\cite{Aad:2015zva,Khachatryan:2014rra} and compare them to the respective 
limits obtained in the EFT approximation. For realistic values of the couplings, 
$g_\chi, g_q\lesssim1$, the LHC provides limits on the messenger mass 
$M_V\lesssim1\,\text{TeV}$. As these are accessible energies at LHC collisions, 
contributions from on-shell messenger production can be large. Hence, limits from 
the simplified model and the EFT can differ significantly as we will discuss in 
Sec.~\ref{sec:LHC}.

As a complementary constraint on the parameter space we consider limits on the 
spin-dependent WIMP-proton scattering from DM annihilation in the Sun.
To this end we re-interpret limits from IceCube~\cite{Aartsen:2012kia} within 
our model, where annihilation into top quark pairs and pairs of mediators are important.
These limits are particularly constraining for large DM masses where the LHC looses 
its sensitivity. We discuss them in Sec.~\ref{sec:IceCube}.
In Sec.~\ref{sec:concl} we conclude.

\section{LHC mono-jet constraints}\label{sec:LHC}

In this work we interpret two searches for mono-jet plus missing transverse 
momentum (MET) signatures performed by ATLAS~\cite{Aad:2015zva} and 
CMS~\cite{Khachatryan:2014rra} at the $8\,$TeV LHC\@. To this end
we performed a Monte Carlo simulation of the signal and imposed the 
search cuts detailed in \cite{Aad:2015zva,Khachatryan:2014rra}. Based 
on the background analysis provided in these references we are thus able 
to set 95\% CL exclusion limits on the parameters of the model. For details 
we refer to~\cite{Heisig:2015ira}.

The considered model has four independent parameters. The DM mass, 
$m_\chi$, the mediator mass, $M_V$, and the couplings of the mediator 
to the DM, $g_\chi$, and the SM quarks $ g_q$. We assume universal 
couplings to all SM quarks and neglect couplings to leptons.
We show our results for various slices of the parameter space where 
we fix the product of the couplings, $g_\chi g_q$ and the mediator width, 
$\Gamma_V$. We choose this parametrization as the cross section for 
DM production directly depends on these parameters. However, not all 
values of $\Gamma_V$ and $g_\chi, g_q$ are consistent within 
this model as we will show below.

For comparison we also derive limits in the EFT approximation
of the considered model. For this we integrate out the messenger to obtain a 
4-fermion contact interaction with an effective coupling $d=g_\chi g_q/M_V^2$. 
Hence, the parameter space reduces to two parameters, $m_\chi$ and $d$. 

\begin{figure}[h!]
\centering
\setlength{\unitlength}{1\textwidth}
\begin{picture}(1.,0.87)
  \put(-0.01,0.48){ 
  \put(0.08,0.025){\includegraphics[width=0.4\textwidth]{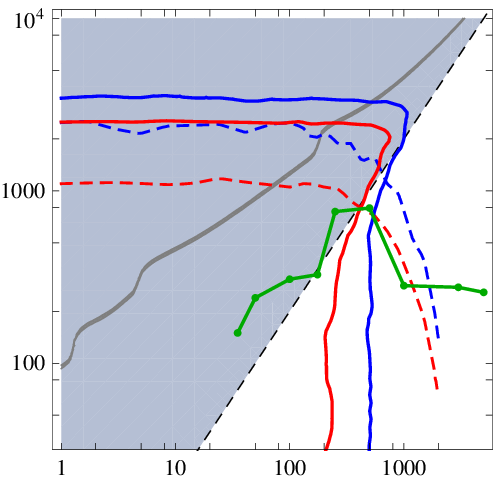}}
  \put(0.123,0.415){\small $\sqrt{g_\chi g_q}=1$, $\Gamma_V=0.01 M_V$}
  \put(0.26,0.0){\footnotesize $m_\chi \,[\text{GeV}]$}
  \put(0.05,0.16){\rotatebox{90}{\footnotesize $M_{V} \,[\text{GeV}]$}}
  \put(0.131,0.15){\rotatebox{44}{\tiny $\uparrow$ $\Omega h ^2 $ too large}}
  \put(0.232,0.06){\rotatebox{57}{\tiny $\uparrow$ no $\Gamma_{\!V}$ sol.}}
  \put(0.414,0.3){\rotatebox{57}{\tiny $M_{\!V}\!=\!2m_\chi$}}
  \put(0.255,0.337){\rotatebox{0}{\tiny \color{blue}  ATLAS mono-jet}}
 \put(0.15,0.28){\rotatebox{0}{\tiny \color{red}  CMS mono-jet}}
  \put(0.433,0.131){\rotatebox{-82}{\tiny  EFT}}
 \put(0.353,0.184){\rotatebox{-92}{\tiny simplified model}}
  \put(0.227,0.175){\rotatebox{0}{\tiny \color{darkgreen}  IceCube}}
  }
 \put(0.48,0.48){ 
  \put(0.08,0.025){\includegraphics[width=0.4\textwidth]{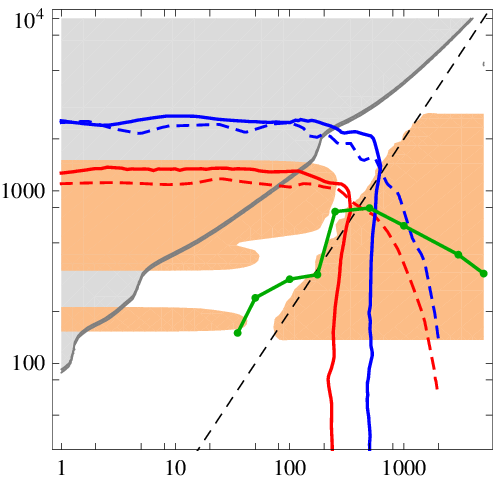}}
  \put(0.123,0.415){\small $\sqrt{g_\chi g_q}=1$, $\Gamma_V=0.5 M_V$}
  \put(0.26,0.0){\footnotesize $m_\chi \,[\text{GeV}]$}
  \put(0.05,0.16){\rotatebox{90}{\footnotesize $M_{V} \,[\text{GeV}]$}}
  \put(0.131,0.15){\rotatebox{44}{\tiny $\uparrow$ $\Omega h ^2 $ too large}}
  \put(0.414,0.3){\rotatebox{57}{\tiny $M_{\!V}\!=\!2m_\chi$}}
 \put(0.02,0.09){
   \put(0.39,0.188){\rotatebox{0}{\tiny \color{darkorange} \;Dijet}}
   \put(0.386,0.17){\rotatebox{0}{\tiny \color{darkorange} (Chala}}
   \put(0.39,0.152){\rotatebox{0}{\tiny \color{darkorange} \;et\,al.)}}
  }
  \put(0.22,0.325){\rotatebox{0}{\tiny \color{blue}  ATLAS mono-jet}}
 \put(0.16,0.283){\rotatebox{0}{\tiny \color{red}  CMS mono-jet}}
  \put(0.433,0.131){\rotatebox{-82}{\tiny  EFT}}
 \put(0.353,0.184){\rotatebox{-92}{\tiny simplified model}}
  \put(0.227,0.175){\rotatebox{0}{\tiny \color{darkgreen}  IceCube}}
}
 \put(-0.01,0.0){ 
  \put(0.08,0.025){\includegraphics[width=0.4\textwidth]{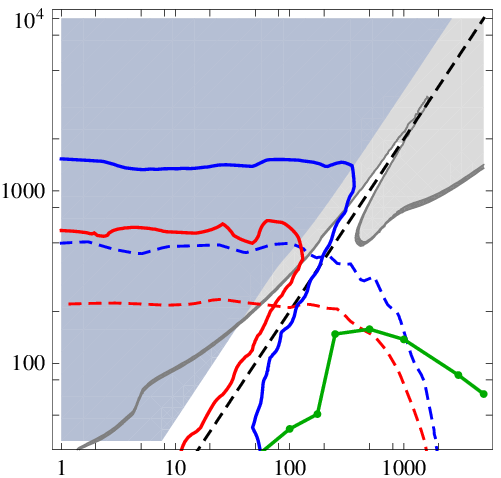}}
  \put(0.123,0.415){\small $\sqrt{g_\chi g_q}=0.2$, $\Gamma_V=0.01 M_V$}
  \put(0.26,0.0){\footnotesize $m_\chi \,[\text{GeV}]$}
  \put(0.05,0.16){\rotatebox{90}{\footnotesize $M_{V} \,[\text{GeV}]$}}
 \put(0.14,0.062){\rotatebox{42}{\tiny $\uparrow$ $\Omega h ^2 $ too large}}
  \put(0.343,0.09){\rotatebox{0}{\tiny \color{darkgreen}  IceCube}}
  \put(0.19,0.29){\rotatebox{0}{\tiny \color{blue}  ATLAS mono-jet}}
 \put(0.17,0.24){\rotatebox{0}{\tiny \color{red}  CMS mono-jet}}
  \put(0.355,0.28){\rotatebox{57}{\tiny $\uparrow$ no $\Gamma_{\!V}$ sol.}}
  \put(0.413,0.295){\rotatebox{57}{\tiny $M_{\!V}\!=\!2m_\chi$}}
  }
 \put(0.48,0.0){ 
  \put(0.08,0.025){\includegraphics[width=0.4\textwidth]{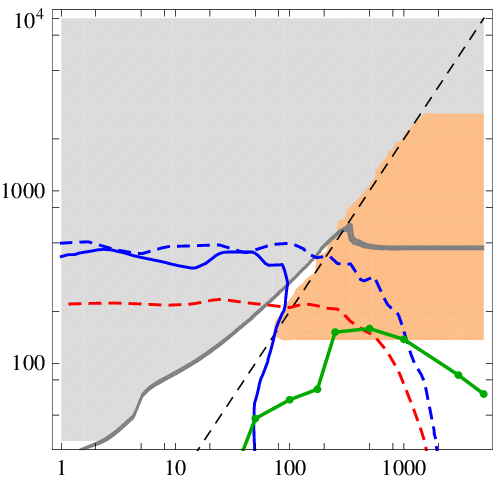}}
  \put(0.123,0.415){\small $\sqrt{g_\chi g_q}=0.2$, $\Gamma_V=0.5 M_V$}
  \put(0.26,0.0){\footnotesize $m_\chi \,[\text{GeV}]$}
  \put(0.05,0.16){\rotatebox{90}{\footnotesize $M_{V} \,[\text{GeV}]$}}
  \put(0.342,0.09){\rotatebox{0}{\tiny \color{darkgreen}  IceCube}}
  \put(0.2,0.224){\rotatebox{0}{\tiny \color{blue}  ATLAS mono-jet}}
 \put(0.15,0.175){\rotatebox{0}{\tiny \color{red}  CMS mono-jet}}
 \put(0.14,0.062){\rotatebox{42}{\tiny $\uparrow$ $\Omega h ^2 $ too large}}
  \put(0.413,0.295){\rotatebox{57}{\tiny $M_{\!V}\!=\!2m_\chi$}}
 \put(0.02,0.08){
   \put(0.39,0.188){\rotatebox{0}{\tiny \color{darkorange} \;Dijet}}
   \put(0.386,0.17){\rotatebox{0}{\tiny \color{darkorange} (Chala}}
   \put(0.39,0.152){\rotatebox{0}{\tiny \color{darkorange} \;et\,al.)}}
  }
  }
\end{picture}
\caption{
Exclusion limits in the $m_\chi$-$M_{V}$ plane in four slices of the considered 
parameter space regarding $\sqrt{g_\chi g_q}$ and $\Gamma_V$~\cite{Heisig:2015ira}.
We show the 95\% CL lower exclusion limits from mono-jet searches at ATLAS (blue lines) 
and CMS (red lines) for the simplified model (solid lines) and the 
EFT approximation (dashed lines). Further, we show 90\% CL lower exclusion limits 
from the IceCube Neutrino Observatory (green lines). 
The dark grey shaded band denotes the region where the thermal relic density 
matches the DM density within $\pm10\%$. In the light-grey shaded
region above it DM is over-produced. 
In the blue shaded region in the left panels no solution exist 
for the individual couplings $g_\chi, g_q$ requiring 
$ M_V,m_\chi,\sqrt{g_\chi g_q}$ and $\Gamma_V$. The 
orange shaded regions are excluded by searches for 
resonances in di-jet signatures taken from Ref.~\cite{Chala:2015ama}.}
\label{fig:mainres}
\end{figure}

In Fig.~\ref{fig:mainres} we show the exclusion limits for the EFT (dashed lines) 
and the simplified model (solid lines) for four slices of the considered parameter 
space. Whilst the EFT limit extents to very high DM masses above a TeV 
the limit from simplified models goes down very drastically for $M_V<2m_\chi$. 
In this region the EFT approximation significantly over-estimates the sensitivity.
Further, also for $M_V\gg m_\chi$ we find significant deviations in the resulting 
limit on $M_V$. This is due to the fact that the limit on $M_V$ placed for 
$\sqrt{g_\chi g_q}\lesssim1$ lies in the region of reachable LHC energies.
This can lead to both an under- or over-estimation of the sensitivity.
On the one hand, contribution from on-shell mediator production could greatly 
enhance the cross section. This is the dominant effect for the parameter slices 
with $\Gamma_V=0.01 M_V$ (left panels of Fig.~\ref{fig:mainres}). The effect 
becomes more pronounced for the smaller coupling, $\sqrt{g_\chi g_q}=0.2$, 
(see lower panels) as the limits are placed at lower $M_V$ where the contribution 
from on-shell mediator production is even larger. In this region EFT under-estimates 
the sensitivity. On the other hand, for very small $M_V$ (below the energy scale 
of the required MET in the considered searches) the cross section is reduced 
and the EFT approximation again over-estimates the limit. This can be seen in 
the case $\sqrt{g_\chi g_q}=0.2$, $\Gamma_V=0.5 M_V$ (lower right panel) 
where the CMS limit for the simplified model completely vanishes whilst the 
EFT would exclude $M_V\gtrsim200\,\text{GeV}$.

As mentioned above not all combinations of $m_\chi$, $M_V$, $\sqrt{g_\chi g_q}$ 
and $\Gamma_V$ are consistent within the model.
In Fig.~\ref{fig:mainres} we marked in blue the regions where no such solution exist.
Note that the region $M_V>2m_\chi$ -- the region where the EFT shows its best 
agreement -- is barely accessible for a reasonably 
small width of $\Gamma_V=0.01 M_V$.

\section{Constraints from DM annihilation in the Sun}\label{sec:IceCube}

If WIMPs scatter in heavy objects like the Sun, they can loose enough energy 
to become gravitationally trapped and accumulate inside the Sun. This leads to a
locally enhanced WIMP density providing significant DM annihilation. Neutrinos
that are produced as primary or secondary products of such annihilations
can escape the Sun and be detected on Earth. On large time-scales an equilibrium 
between the capturing and annihilation will be be reached. In this case, a limit on the 
flux of neutrinos produced in WIMP annihilations can be translated into a limit on the 
scattering cross section of WIMPs inside the sun. As the Sun contains large amounts 
of hydrogen it provides sensitivity to spin-dependent WIMP-proton scattering.
 
We use data from the IceCube Neutrino Observatory~\cite{Aartsen:2012kia} which is 
interpreted in two benchmark scenarios according to DM annihilating 100\% into $b\bar b$ 
or $WW$. In most of the parameter space of our model, annihilation into $b\bar b$,
$t\bar t$ or $VV$ dominates. Therefore we re-interpret the limits on the spin-dependent
WIMP-proton cross section from Ref.~\cite{Aartsen:2012kia} and estimate a limit for 100\% 
annihilation into $t\bar t$ and $VV$ as a function of $m_\chi$ (and $M_V$)~\cite{Heisig:2015ira}.
We then apply the limit to our model parameter space taking into account the respective 
contribution to annihilation of the channels $b\bar b$, $t\bar t$ and $VV$. 
The resulting limits are shown in Fig.~\ref{fig:mainres} (green lines). In the region of
large $m_\chi$ where LHC searches loose sensitivity, the limits from IceCube are able to 
exclude mediator masses up to $M_V\simeq 200\,\text{GeV}\; (1\,\text{TeV} )$ for
$\sqrt{g_\chi g_q}=0.2\; (1)$. 

Although the capture of WIMPs in the Sun is well described by EFT, 
the annihilation process is in general not. As a consequence of the different solutions
for the individual couplings $g_\chi$ and $g_q$ for different mediator widths
chosen in our parameter slices the relative contributions of the annihilation channels 
$b \bar b$, $t \bar t$ and $VV$ can differ drastically. 
Since limits for annihilation into $VV$ are much weaker than for $t \bar t$,
a large $VV$ contribution can significantly weaken the IceCube limit. This 
is the case for $\sqrt{g_\chi g_q}=1$, $\Gamma_V=0.01 M_V$ and 
$m_\chi\ge 1\,$TeV, where annihilation into $VV$ is particularly important.

\section{Conclusion} \label{sec:concl}

We have considered a model with a vanishing spin-independent WIMP-nucleon cross section 
and set new limits on the model parameter space from LHC mono-jet searches as well as 
IceCube.  Using LHC limits for $\sqrt{g_\chi g_q}= 1$ we can exclude mediator masses up to
around 3\,TeV for $m_\chi\lesssim1\,\text{TeV}$ while for  $\sqrt{g_\chi g_q}= 0.2$ we exclude
$M_V$ in the range of 500\,GeV to 1.5\,TeV with a strong dependence on the mediator width.
We have compared these limits to the ones obtained in the EFT and found that those can both
over- or under-estimate the sensitivity depending on the corner in parameter space. 
Limits from IceCube are complementary probing the region of large $m_\chi$ where the LHC 
is not sensitive at all reaching up to $M_V\simeq1\,\text{TeV}$ for $\sqrt{g_\chi g_q}= 1$.


\begin{footnotesize}

\end{footnotesize}


\end{document}